# Dante: An Open Source Model Pre-Training and Fine-Tuning Tool for the Dafne Federated Framework for Medical Image Segmentation

Giuseppe Timpano [a,b,*], Dibya Kumari [b], Maria Eugenia Caligiuri [a,c], Francesco Santini [b,d]

[a] Department of Surgical and Medical Sciences, Magna Graecia University, Catanzaro 88100, Italy
[b] Basel Muscle MRI, Department of Biomedical Engineering, University of Basel, 4123 Allschwil, Switzerland
[c] Neuroscience Research Center, Magna Graecia University, Catanzaro 88100, Italy
[d] Department of Radiology, University Hospital Basel, 4031 Basel, Switzerland

* Corresponding author: Giuseppe Timpano (giuseppe.timpano@unicz.it)

## Abstract

Adapting pre-trained deep learning segmentation models to new clinical domains is a persistent challenge in medical image analysis, particularly when annotated data at the target site are scarce. Parameter-efficient fine-tuning strategies offer a principled solution by selectively updating a controlled subset of model parameters, preserving previously acquired representations while reducing the risk of overfitting on small datasets.

This paper introduces DAfNe TrainEr (Dante), an open-source module integrating with the Dafne federated segmentation ecosystem as a dedicated training and fine-tuning backend. Dante supports training from scratch with automatic architecture configuration, configurable layer freezing schedules, and Low-Rank Adaptation (LoRA) extended to N-dimensional convolutional layers through channel-wise factorization. To validate the module, Gradual Unfreezing (GU) and LoRA are assessed across realistic cross-domain MRI transfer scenarios covering abdominal organ segmentation and brain white matter lesion segmentation, under full-data and few-shot conditions.

GU reduced the epochs required to reach 85% of peak performance by up to 63.6% compared to training from scratch, while LoRA achieved Dice Similarity Coefficients up to 0.957 in data-rich scenarios. Both strategies outperformed the baseline across all tested domains, with gains amplified by richer pre-training datasets.

These results suggest that Dante can support the adaptation of medical image segmentation models across the evaluated anatomical and imaging domains when annotated target data are limited.

Dante code is available at https://github.com/dafne-imaging/dafne-torch-trainer while Dafne ecosystem project is available at https://github.com/dafne-imaging.

# 1. Introduction

In recent years, automatic segmentation techniques in the medical field have exhibited human-like performance on real clinical cases, significantly reducing the reliance on human interpretation. Automated segmentation of anatomical structures from medical images is an important step for quantitative clinical analysis to support volumetric organ assessment, surgical planning, radiotherapy contouring, and the extraction of imaging biomarkers for longitudinal follow-up. Deep learning architectures, especially those based on the U-Net encoder-decoder paradigm and its variants like Vision Transformers (ViT) and hybrid models such as UNetR and Swin-Unet, as well as large-scale foundation models [1,2], have shown state-of-the-art performance in public benchmarks, achieving Dice similarity coefficients that approach inter-expert agreement for various anatomical structures in CT and MRI modalities [3]. These successful segmentation models are, however, often trained on large datasets, comprising thousands of cases [4,5]. On the other hand, clinical reality is often marked by a persistent scarcity of data. Acquiring and validating annotations from expert radiologists can be costly and impractical, which hinders the efficacy of ad-hoc models trained on local data.

A significant challenge to the practical implementation of these models is domain shift: the systematic discrepancy between the data distribution encountered during training and the one observed during inference. In medical imaging, this discrepancy arises from differences in scanner hardware, acquisition protocols, reconstruction parameters, patient demographics, and annotation conventions across clinical sites or among radiologists performing manual segmentation. Generally, a model pre-trained on an institutional dataset, particularly for MRI images (which are highly susceptible to inter-patient variability and inter-operator differences in segmentation), is unlikely to maintain its performance when applied to data from a different site or protocol without some form of adaptation [6,7]. This issue represents a significant barrier to the widespread implementation of high-performance segmentation models in diverse clinical environments.

Transfer learning from a pre-trained model helps mitigate this constraint. Therefore, selective and parameter-efficient fine-tuning strategies that adapt only specific subsets of model parameters are particularly valuable for real-world applications.

In this context, Dafne (Deep Anatomical Federated Network) [8] is an open-source, cross-platform client-server system designed for collaborative medical image segmentation through federated incremental learning. It hosts a global deep learning model on a central server, where multiple nodes download the model and enable end users to refine predictions through an integrated interface. This process triggers a local incremental update that is sent back to the server and incorporated into the global model. As a result, Dafne facilitates continuous, privacy-preserving improvements to segmentation models without centralizing raw patient data. It has shown statistically significant performance enhancements in real-world applications [8].

Dafne relies on the continuous improvement of an existing model by the usage and refinement of the output masks by the users. To this end, it is necessary to have an initial pretrained model that can provide a reasonable initial performance as a starting point for manual refinement.

This paper introduces DAfNe TrainEr (Dante), an open-source module that serves as an automated training and fine-tuning backend within the Dafne ecosystem. Alongside training from scratch with automatic architecture adaptation to the input data, it features two complementary selective fine-tuning strategies for segmentation models: Layers Freezing (with also Gradual Unfreezing strategy) and LoRA (Low-Rank Adaptation) applied to convolutional and transposed-convolutional layers of segmentation models. To empirically validate this module, both strategies are evaluated across various realistic transfer scenarios utilizing publicly available MRI datasets. These datasets cover different anatomical domains and MRI imaging protocols. A detailed description of the experimental design and specific scenarios can be found in Section 3.

Transfer learning from pre-trained models is the standard strategy for addressing data sparsity and domain shift in medical image segmentation. The main theoretical basis is the concept that early convolutional layers encode general, transferable features, while deeper layers encode task-specific representations [7]. This layer specialization has led to studies on CNNs showing that fine-tuned CNNs consistently outperform models trained from scratch in medical imaging tasks, particularly with sparse target domain annotations. Additionally, the depth of adaptation should correspond to the extent of domain shift [7], and encoder filters change minimally during fine-tuning in cross-modal and cross-organ transfer scenarios [7]. This supports the rationale of progressive layer unfreezing to preserve low-level representations while allowing higher layers to adapt.

## 1.1. Related Works

The application of Parameter-Efficient Fine-Tuning (PEFT) to medical image segmentation has been driven largely by large vision foundation models. Hu and colleagues [9] introduced LoRA, which decomposes weight updates into low-rank matrices and has become the dominant PEFT method across vision and language tasks. In medical imaging, LoRA has been applied primarily to Transformer-based architectures such as SAM [10], with Conv-LoRA [11] introducing spatial inductive biases into the low-rank matrices to better handle dense prediction tasks. The most comprehensive evaluation of PEFT in medical imaging to date, by Dutt and colleagues [12], tested 17 algorithms across six datasets and found that PEFT outperforms full fine-tuning in low-data regimes — but identified a systematic gap: convolutional encoder-decoder networks remain largely unstudied. Silva-Rodríguez and colleagues [13] partially addressed this with the FSEFT framework for volumetric organ segmentation, demonstrating that LoRA-adapted models with 1–5 annotated volumes match fully fine-tuned models trained on 30+ volumes.

No existing tool provides a configurable, ready-to-deploy fine-tuning backend for federated medical image segmentation systems. Automated segmentation tools such as TotalSegmentator [5] and MRISegmenter [14] offer robust pre-trained models but no mechanism for local adaptation. The Dafne system [8] enables federated incremental learning but cannot, by itself, extend the existing provided models to new segmentation problems, and necessitates an initial pretrained model for its manually assisted learning approach to be viable.

# 2. Material and Methods

Dante is a training module designed to integrate with the Dafne ecosystem, enhancing its segmentation pipeline with structured fine-tuning support. Users have the option to either train a segmentation model from scratch or adapt an existing one—such as the global model from the Dafne server or a locally trained checkpoint—using a limited set of annotated local data.

The primary objective is not to produce a polished, production-ready model but to provide each federation node with a simple method to customize an existing model to fit its specific data needs. Each adaptation begins with a model that already possesses significant knowledge about the relevant anatomy, acquired either from a pre-training dataset or through federated contributions from various nodes. The adaptation process should focus on updating the most relevant parameters for the local data while preserving the existing knowledge, ensuring that the model's prior learning is retained and not discarded.

Dante provides three base architectures for developing segmentation models from scratch: 2D U-Net, 3D U-Net, and DynUNet. Each architecture allows the user to customize network parameters, such as kernel size, stride size, and network depth, based on the characteristics of the input data. The DynUNet architecture is based on the automatic configuration of training parameters and network, following the self-configuration principle of nnU-Net [15], enabling it to adjust to the specific voxel spacing, image size, and modality of the target dataset without manual hyperparameter tuning. Furthermore, fine-tuning strategies like Layers Freezing and LoRA are available for all three architectures, allowing users to adapt any pre-trained model to new local data, regardless of the training mode originally used.

Figure 1 illustrates the overall architecture of Dante and its two operational modes. Both modes utilize a common input format and generate a model file compatible with the Dafne ecosystem. The key difference lies in whether the network parameters are initialized from scratch or inherited from a pre-trained checkpoint.

## 2.1. Data preparation

All three architectures utilize a shared data preparation pipeline. Input data is retrieved from numpy bundle files in a Dafne-specific format (*npz bundle files)* containing the image volume, segmentation masks, and original voxel resolution (Figure 1). Once loaded, the volume undergoes three sequential operations: foreground cropping, resampling to the target spacing, and intensity normalization.

Resampling addresses anisotropic acquisitions, ensuring spatial coherence along the dominant resolution axis. Intensity normalization is applied to non-zero voxels on a channel-by-channel basis, preventing background regions from influencing the normalization statistics.

For DynUNet network hyperparameters and batch size are automatically determined based on a dataset fingerprint that summarizes the median voxel spacing and median image shape of the training set. The patch size is also constrained by the available GPU memory. In contrast, for 3D and 2D U-Nets, some parameters such as the patch size (for the sliding window training and inference [16]) are fixed, while others, such as the batch size and network hyperparameters, can be manually defined by the user (see Table 1 for more details). Augmentation operations, manually selectable by the user, include random inversions, rotations, zooms, and the addition of Gaussian noise. In 3D model training, individual volumes are divided into patches (automatically or manually defined) using a 3:1 positive-to-negative labeling ratio, ensuring that each batch contains a sufficient proportion of annotated foreground regions. For the 2D U-Net, patch extraction is replaced with slice-level selection, retaining only slices that contain at least one positive annotation along with their immediate surroundings (Table 1).

| | DynUnet | U-Net 3D | U-Net 2D |
|---|---|---|---|
| **Training and inference patch size** | automatic (VRAM-aware) (Scratch)<br>Inherited (FT Mode) | (16, 96, 96) (Scratch)<br>Inherited (FT Mode) | N/A (slice-based) |
| **Network hyperparameters** (kernels, strides, levels…) | Dataset fingerprint (Scratch) / Inherited (FT Mode) | User-defined | User-defined |
| **Batch size** | Dataset fingerprint (Scratch) / Inherited (FT Mode) | User-defined | User-defined |
| **Target spacing** | Dataset fingerprint (Scratch)<br>Inherited (FT Mode) | Dataset fingerprint (Scratch)<br>Inherited (FT Mode) | Dataset fingerprint (Scratch)<br>Inherited (FT Mode) |

**Table 1**: Comparison of configuration parameters across the three Dante training architectures. Automatic: determined from dataset fingerprint and available GPU memory. Manual: defined by the user prior to training. Scratch: Model training from scratch. FT mode: Fine-tuning of an existing model.

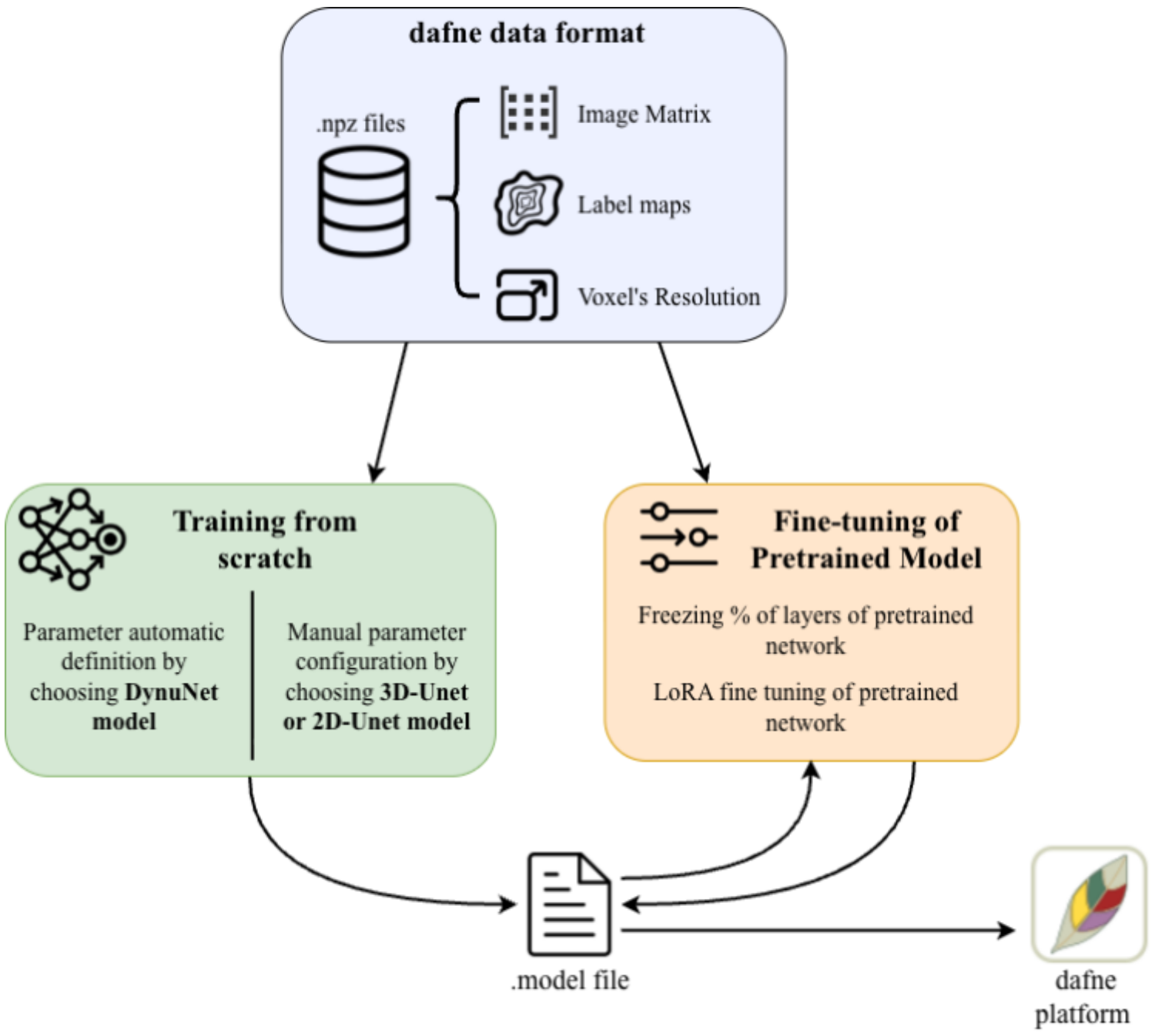

**Figure 1**: Overview of the Dante Workflow: Dante accepts input data in Dafne .npz bundle format, which includes image volumes, label maps, and voxel resolution. It offers two operational modes: i) training from scratch, using either a DynUNet with automatic parameter configuration or a 2D/3D U-Net with manual configuration; ii) fine-tuning a pre-trained model through freezing layers or LoRA. The resulting model is saved as a .model file, ensuring direct compatibility with the Dafne inference pipeline.

## 2.2. Fine Tuning strategies

Dante provides two independent axes of control for fine-tuning a pre-trained model. The first governs the freezing schedule of the network layers. The second controls whether low-rank LoRA convolutional adaptation matrices are injected into the convolutional layers.

### Freezing schedule.

Two strategies are available: static partial freezing, which locks a fixed proportion of layers, and Gradual Unfreezing (GU), which progressively unlocks layers during training. At the beginning of training, only the output-proximal layers were trainable, while the earlier feature extraction layers were kept frozen to maintain the pre-trained knowledge. As training continued, the network was gradually unfrozen from the bottom up—starting with the output layers and moving toward the input layers—at fixed intervals of 10% of the total training epochs. By the time the training reached the 75th percentile, the entire network had become trainable. All normalization layers (Batch Normalization and Instance Normalization layers) remained active and trainable throughout the process to ensure statistical stability during domain adaptation. The learning rate can either follow a cosine annealing schedule or remain constant, depending on user preference.

### LoRA conv modules

Independent of the freezing schedule, Low-Rank Adaptation can be enabled to inject trainable low-rank matrices into the convolutional layers while keeping original weights frozen. The rank $r$ and scaling factor $\alpha$ are configurable hyperparameters, allowing control over the adaptation capacity and the weighting between pre-trained and adapted representations.
A Low-Rank Adaptation (LoRA) for generic N-dimensional convolutional layers was proposed, extending the original formulation by Hu et al. [9,11] to the convolutional weight tensor through channel-wise factorization. The underlying assumption is that parameter redundancy in a convolutional model primarily arises from the interactions between the input and output channels, rather than from the spatial structure of the kernel.

The integration of the Low-Rank Adaptation (LoRA) methodology into the convolutional layers consists into decomposing the weight update ($\Delta W$) matrix into the multiplication of two low-rank matrices, $A$ and $B$. This approach leverages the linearity of the convolution operator, enabling to separate the computation of the pre-trained (frozen) weights from that of the additive adaptation (Figure 2). Based on the concept of linearity of the convolution operation (eq. (1)):

$$y = Conv(x, W) + Conv(x, \Delta W_k) \tag{1}$$

$$\Delta W_k = \frac{\alpha}{r}(A_k \cdot B_k) \tag{2}$$

Instead of updating all the elements of the matrix $\Delta W \in \mathbb{R}^{C_{out} \times C_{in} \times K}$, training is performed on the reduced number of elements of the matrices $A_k \in \mathbb{R}^{k1 \times \cdots k_n \times C_{out} \times r}$ and $B_k \in \mathbb{R}^{k_1 \times \cdots k_n \times r \times C_{in}}$ (eq.

(2)). To ensure the injection of convolutional LoRA modules in all convolution and transposed convolution operations, the implemented LoRA module adapts to both scenarios by permuting the resulting $(\Delta W)$ to match the expected weight layouts: $([C_{out}, C_{in}, \ldots])$ for standard convolutional layers and $([C_{in}, C_{out}, \ldots])$ for transposed layers.

### Optimizer and training hyperparameters

For most experiments, datasets were partitioned into training and validation sets using an approximately 80/20 patient-wise split. The ISBI 2015 dataset comprised 21 longitudinal FLAIR acquisitions from five patients, with four to five time points per patient. Three patients were assigned to training and two to validation. All acquisitions from each patient were kept in the same partition. Models were trained from scratch of all three available architectures (2D and 3D UNets, and DynUNet) with fixed augmentation (random 90° rotation, Gaussian random noise and random zoom). Models were trained using the AdamW optimizer and with a weight decay of $\lambda = 10^{-4}$ to mitigate overfitting [17]. Furthermore, a cosine annealing learning rate schedule was adopted to smoothly decaying the learning rate from its initial value to a minimum of $\eta_{\min} = 10^{-6}$ over the total number of training epochs $T_{\max}$ (set to 200 for all experiments with early stopping patience of 20 epochs). For fine-tuning of the pre-trained models the modality remained the same, but the augmentation involved only random noise and random zoom on the training dataset. All tests were executed in Python 3.10.20 environment using the PyTorch 2.8.0 framework on a single NVIDIA RTX 5000 Ada GPU (32 GB VRAM).

## 2.3. Study Dataset

The experimental validation utilized five open-source MRI datasets, divided into two anatomical domains: abdominal MRI datasets and brain MRI datasets. The abdominal MRI group included AMOS, CHAOS, and the NIH MRISegmenter dataset, while the brain MRI group comprised MSLesSeg and the ISBI 2015 Longitudinal MS Lesion Segmentation dataset.

In the abdominal MRI datasets, AMOS [18] features 100 MRI scans with voxel-level annotations for 15 abdominal organs. This dataset encompasses multi-center, multi-vendor, and multi-disease cases collected from eight different MRI scanners, providing a diverse source domain for multi-organ segmentation. Only axial MRI scans were considered for this study. For the AMOS experiments, 42 MRI cases were selected for source-domain pre-training (34 training and eight validation), while a smaller subset of 10 cases was used for target-domain fine-tuning (eight training and two validation). These subsets were not mutually exclusive across different experimental scenarios; however, patient independence between training and validation was maintained within each individual experiment. CHAOS [19] contains abdominal MRI data from 20 healthy volunteers, acquired using two pulse sequences: Dual-Echo T1-weighted Gradient Echo (T1-DUAL) and T2-weighted Spectral Pre-saturation with Inversion Recovery (T2-SPIR). It includes manual annotations for four abdominal structures: liver, spleen, right kidney, and left kidney. The NIH MRISegmenter dataset [14] consists of images from 195 patients imaged at the NIH Clinical Center, utilizing axial precontrast and multiphase contrast-enhanced T1-weighted sequences; only the portal venous phase was used in this work. Seven anatomical structures across the selected series were considered for pre-training: liver, spleen, right kidney, left kidney, stomach, inferior vena cava, and pancreas.

The brain MRI datasets were employed to evaluate Dante in the context of lesion segmentation. MSLesSeg [20] contains brain MRI scans from multiple centers, acquired using various 1.5T protocols, with expert annotations of white matter lesions based on T1, T2, and T2-FLAIR sequences. The ISBI 2015 Longitudinal MS Lesion Segmentation dataset [21] includes 21 longitudinal MRI acquisitions from five subjects with four to five time points per patient, acquired on a 3T scanner at a single centre. For both datasets, the analysis concentrated on T2-FLAIR sequences, creating a scenario of adaptation across centres and field strengths under limited data conditions.

From these datasets, five transfer-learning scenarios were established. In the first abdominal scenario, pre-training was conducted on AMOS, followed by fine-tuning on CHAOS, with evaluation limited to the four structures consistently annotated in both datasets: liver, spleen, right kidney, and left kidney. The second scenario reversed the transfer direction, pre-training on CHAOS and fine-tuning on a subset of 10 patients from AMOS. In the third scenario, the model was pre-trained on the NIH MRISegmenter dataset and fine-tuned on the same 10-patient subset from AMOS, again focusing on the four structures shared with CHAOS. The fourth scenario involved pre-training on NIH using labels that differed from the AMOS target labels to evaluate transfer learning across different segmentation targets. Finally, in the brain MRI scenario, pre-training was performed on MSLesSeg, followed by fine-tuning on ISBI 2015 to assess adaptation for multiple sclerosis lesion segmentation in a context of extreme data scarcity. A summary of the datasets, acquisition characteristics, labels, and experimental roles is presented in Table 2.

## 2.4. Training experiments

The performance of the system was tested under the following conditions: training from scratch of all three architectures on the CHAOS dataset, and multi-organ transfer learning by using the various combinations of GU and LoRA strategies starting from base models pretrained on AMOS, CHAOS, NIH, and MSLesSeg datasets, respectively, using the DynUNet architecture. When AMOS was used for fine tuning, a subset of 10 patients was used, to simulate a realistic scenario in which fine tuning is performed on a smaller dataset than the one originally used for training.

All the experiments above were performed to evaluate domain shift on the same segmentation problem (same output labels). One additional experiment was also performed to test the validity of the transfer learning to a different segmentation problem, by using different labels between the initial pretraining and the second fine-tuning step, from dataset NIH (with labels stomach, aorta, inferior vena cava, pancreas) to dataset AMOS (with CHAOS' labels).

More information about the experimental setup, including pre-training sources, target adaptation tasks, and the anatomical structures segmented in each experimental condition, is available in Table 3.

## 2.5. Evaluation metrics and data analysis

Trends, stability, and convergence speed by measuring the average Dice Similarity Coefficient (DSC) across all segmented labels, considering the limited validation set, were analysed: the primary goal is to validate Dante and assess how its fine-tuning strategies can mitigate the performance penalty caused by domain shift. DSC after completion of the training was considered as the primary endpoint. For transfer learning, the number of epochs needed to reach the peak DSC performance, and the number of epochs needed to reach 85% of the maximum performance were considered as indicators of convergence speed, to compare the various fine-tuning algorithms. The experimental evaluation was conducted using patient-wise training and validation partitions, without a separate independent test set. During training, the DSC was monitored on the validation set, and the checkpoint achieving the highest validation DSC was selected for reporting. Early stopping was applied after a predefined number of consecutive epochs without improvement in validation performance. Therefore, all reported DSC values represent validation performance rather than independent test-set estimates.

| Dataset | Subjects | Modality / Sequence | Scanner / Field Strength | Labels | Role in Experiments |
|---|---|---|---|---|---|
| AMOS ([18]) | 42 (axial MRI only) | MRI — multi-sequence | Multi-centre, multi-vendor, 8 scanners | 15 abdominal organs | Pre-training · Scenario 1 (AMOS→CHAOS, full data) |
| CHAOS ([19]) | 20 | MRI — T1-DUAL, T2-SPIR | Single-centre | Liver, spleen, left kidney, right kidney (4 labels) | Fine-tuning · Scenario 1 (AMOS→CHAOS)<br><br>Pre-training · Scenario 2 (CHAOS→AMOS, N=10)<br><br>Training of 3D and 2D UNet |
| NIH Dataset (MRISegmenter) ([14]) | 195 (venous phase only) | MRI — T1w portal venous | Single-centre, NIH Clinical Center | Liver, spleen, R. kidney, L. kidney, stomach, inferior vena cava, pancreas (7) | Pre-training · Scenario 3 (NIH→AMOS, N=10).<br><br>Pre-training · Scenario 6 (NIH→AMOS, N=10). |
| MSLesSeg ([20]) | 75 | MRI — FLAIR | Multi-centre, 1.5T | White matter lesions (1 label) | Pre-training · Scenario 4 (MS cross-centre) |
| ISBI 2015 ([21]) | 5 | MRI — FLAIR | Single-centre, 3T | White matter lesions (1 label) | Fine-tuning · Scenario 4 (MS cross-centre, extreme few-shot) |

**Table 2.** Summary of Datasets Used in the Experimental Validation of Dante. Subject counts reflect the subsets utilized in these studies. AMOS provides annotations for 15 abdominal organs; however, evaluation in Scenario 2 and 3 was limited to the 4 structures for AMOS dataset: liver, spleen, left kidney, and right kidney.

# 3. Results

To evaluate the adaptation of the pre-trained model in a domain transfer scenario, we conducted several fine-tuning experiments based on GU and LoRA using a small target dataset. An example of Dante during training (or fine-tuning) is shown in Figure 3.

## 3.1. Scratch Training Benchmarks

To evaluate the manually defined UNet models (both 2D and 3D versions), a simple training session was conducted from scratch using the CHAOS dataset. The results indicate that UNet-2D significantly outperformed UNet-3D in terms of generalization, achieving an average Dice Similarity Coefficient (DSC) of 0.725, compared to UNet-3D's 0.548. DynUNet outperformed both UNet-2D and UNet-3D with a DSC of 0.84 on the same dataset and was thus used as a basis for the fine-tuning experiments.

## 3.2. Multi-Organ Transfer Learning Results

First, the domain shift from AMOS to CHAOS was analyzed (Table 3, Multi-Organ Transfer (A)). The baseline CHAOS model achieved an average Dice Similarity Coefficient (DSC) of 0.84 after 171 epochs, while LoRA ($r = 16$) reached a higher DSC of 0.8726 in only 64 epochs, indicating that the small CHAOS dataset did not provide enough morphological diversity to effectively train the network without pre-training. GU reached 85% of its peak validation DSC in 16 epochs, compared with 44 epochs for the scratch baseline to reach 85% of its own peak. This represents a 63.6% reduction in the number of epochs required to reach the threshold. At the organ level, the liver showed the greatest improvement, with fine-tuning strategies utilizing pre-trained anatomical priors to enhance performance (92.61% vs. 83.08% DSC). Although the baseline model retained a slight edge in segmenting the right kidney, the overall stability and accelerated dynamics of the Dante-adapted models demonstrate their effectiveness for data-sparse scenarios (see Table S1 in supplementary material for more details).

Furthermore, to evaluate Dante's performance in a more diverse domain, two small-sample experiments using a subset of 10 patients from the AMOS dataset for fine-tuning were conducted. The only difference between the experiments was the pre-training source. In the first experiment (Table 3, Multi-Organ Transfer (B)), model was pretrained using the CHAOS dataset, while the second experiment was pretrained starting from the NIH dataset (Table 3, Multi-Organ Transfer (C)), which provides a richer anatomical prior. In both experiments, the segmentation task focused on the four shared structures: liver, spleen, and bilateral kidneys.

Dante's fine-tuning strategies consistently outperform the baseline model, regardless of the pre-training source. Both transfer scenarios used the same 10-patient AMOS target subset with the same patient-wise split (8 training and 2 validation patients). A DynUNet trained from scratch on this split served as the reference for both comparisons and achieved a peak average validation DSC of 0.7848, reaching 85% of its peak at epoch 41. With CHAOS pre-training, LoRA (r = 32) reached a peak DSC of 0.8864, and GU reached 85% of its own peak in 6 epochs (Table 4).

When pre-training on the NIH dataset, all fine-tuning strategies produced notably higher DSC values: LoRA (r = 32) achieved 0.957, and GU reached 0.956. GU reached 85% of its own peak in 14 epochs with NIH pre-training. Validation and dice loss curves demonstrate that the NIH pre-training condition leads to significantly smoother and faster convergence curves than CHAOS pre-training, which aligns with the higher final DSC values reported earlier (see Figure S1 in supplementary materials for more details).

By pretraining the model on the NIH dataset and fine-tuning it on the AMOS dataset, while ensuring that labels from both datasets remain mutually exclusive, the results indicate that GU achieves an exceptional balance between accuracy and convergence speed in the new AMOS domain, with a DSC of 0.9318. This performance surpasses that of the LoRA fine-tuning, which produced DSC scores of 0.8860, 0.8990, and 0.9058 for r = 8, 16, and 32, respectively. However, the DSC performance is lower than that of corresponding experiments conducted with non-mutually exclusive labels during pretraining on the NIH dataset. The model's extensive pretraining on segmented data for the stomach, pancreas, aorta, and inferior vena cava enabled it to effectively segment the spleen from the earliest epochs (1, 3, 2, and 4 for GU and LoRA with r = 8, 16, and 32, respectively). In contrast, the model faced greater challenges in recognizing the liver, achieving its peak performance at epochs 97, 107, 58, and 133 for GU and LoRA with r = 8, 16, and 32, respectively.

## 3.3. MS Brain Lesion Adaptation Analysis

In the MS lesion segmentation scenario (Table 3, MS Brain Lesion Adaptation), GU reached 85% of its peak validation DSC at epoch 1 and peaked at 0.849 at epoch 32. The scratch baseline reached 85% of its peak at epoch 17 and peaked at 0.818 at epoch 61. Both LoRA configurations reached 85% of their respective peaks at epoch 2. LoRA (r = 16) peaked at 0.840 at epoch 6, and LoRA (r = 32) peaked at 0.848 at epoch 5 (see Table S2 in supplementary materials).

# 4. Discussion

In this work, we tested different experimental scenarios to assess the ability of Dante's fine-tuning strategies for adapting pre-trained segmentation models to new domains with limited annotated data.
First, we assessed the models' capabilities to learn from data by conducting simple training from scratch on UNet-2D, UNet-3D (with manually defined hyperparameters), and finally, DynUnet. A performance gap was observed between the manually defined models of UNet-2D and UNet-3D due to the challenges associated with volumetric training in data-sparse environments. The 3D U-Net processes sub-volumes (patches) of size (16, 96, 96), a limitation imposed by GPU memory constraints, which restricts the network's effective receptive field. Consequently, it captures less global contextual information than the 2D approach, which analyzes entire axial slices. This finding underscores the advantages of using the DynUNet module within Dante for processing 3D volumes. Unlike conventional 3D U-Net, which relies on manually defined parameters, DynUNet automatically optimizes the patch size based on available VRAM and batch size. By fine-tuning these spatial parameters, DynUNet can capture a broader anatomical context during training than UNet-3D, leading to more accurate contours and improved performance.
In terms of experiments on DynUnet fine-tuning, two complementary strategies were evaluated: GU, which progressively unfreezes the network from decoder to encoder, and LoRA, which updates the convolutional weight matrices via low-rank decomposition while keeping the original weights frozen.
Several key points deserve attention. First, GU demonstrates the ability to converge rapidly, even with the encoder layers frozen during the early training epochs. In experiments involving multiple sclerosis lesion datasets, GU reached 85% of its maximum validation DSC after just one epoch, compared with 17 epochs for the scratch baseline. However, the actual peak validation performance was achieved at epoch 32 for GU and epoch 61 for the baseline, indicating faster early convergence rather than immediate attainment of peak performance. In the AMOS-to-CHAOS abdominal organ segmentation experiment, GU reached 85% of its peak validation DSC in 16 epochs and the scratch baseline reached 85% of its own peak in 44 epochs, a 63.6% reduction in epochs to the threshold. By keeping the lower layers frozen during the initial training stages, GU enables the network to adapt progressively rather than reshuffling all learned features at once. The baseline model lacks this initial advantage and requires significantly more epochs to achieve comparable performance. In few-shot scenarios, where target data is limited, this delay increases the risk of overfitting before the model can converge.
Second, pre-training dataset quality acts as a strong modulator of final performance, independently of the fine-tuning strategy applied. When fine-tuning on the AMOS subset (N=10), pre-training on the NIH dataset (providing 195 patients, 7 annotated structures, and a high-quality T1-weighted protocol) yielded DSC values of 0.955–0.957 across all Dante strategies, compared to 0.873–0.886 when pre-training on the smaller CHAOS dataset. This finding aligns with the role of domain-aligned pre-training as a strong initialization prior and has direct practical implications for federated deployment: the richer and more representative the global model distributed, the more effective the local adaptation will be, even with very few target annotations. When the source and target datasets have mutually exclusive labels, the results show that the Gradual Unfreezing (GU) strategy achieves a better balance between accuracy and convergence speed in the task-shift regime. This improvement over LoRA (with r=32 and a DSC of 0.9058) comes from the greater parametric freedom that GU provides during the initial fine-tuning phases, which helps restructure the decoder to manage mutually exclusive classes. Additionally, the organ-specific analysis reveals varied learning dynamics: the spleen benefits from near-instantaneous feature transfer during the first four epochs, while the liver requires a much longer adaptation phase, spanning epochs 58 to 133. This highlights how anatomical proximity during pre-training, such as with the pancreas and stomach, directly influences the speed of feature reuse in the target domain.
Third, the consistent performance of Dante's fine-tuning strategies across two clinically distinct domains—abdominal organ segmentation using multi-sequence MRI and white matter lesion segmentation from FLAIR brain MRI—validates the module's domain-agnostic design. In both cases, GU and LoRA outperformed the baseline without any task-specific configuration, indicating that the module can be deployed as-is in federated nodes operating on different anatomical targets and imaging protocols, without the need for expert tuning for each new application. In this case, results demonstrate that the advantages of pre-trained representations extend beyond multi-organ segmentation tasks. Even in a single-structure pathology detection task that experiences significant cross-center and cross-field-strength domain shifts, Dante's fine-tuning strategies facilitate quick and stable adaptation using only a small number of annotated cases.
Furthermore, our experiments revealed a difference in generalization performance between the transfer scenarios of AMOS to CHAOS and CHAOS to AMOS. Models pre-trained on AMOS adapted better when fine-tuned on CHAOS due to AMOS being a large-scale, multi-center dataset with a wide range of morphological and pathological diversity across 100 scans, while CHAOS includes only 20 healthy volunteers from a single center. Consequently, the AMOS-trained model developed more robust feature representations that transferred easily to the simpler domain of healthy anatomy in CHAOS. In contrast, the CHAOS-trained model struggled with the greater inter-patient variability and scanner discrepancies in the AMOS dataset.

Despite the results obtained in the proposed experimental sets, the study has a few limitations.
First, the experimental validation relies on a limited number of datasets and transfer scenarios chosen to represent the two main implementation conditions in the Dafne federation—full data domain transfer and extreme few-shot adaptation—but it is not exhaustive. The generalizability of the observed performance trends to other anatomical regions, imaging modalities, and clinical sites requires further investigation, and additional validation on a broader range of datasets should be considered. Second, the comparisons in this work are limited to Dante's fine-tuning strategies and do not include direct numerical comparisons with other PEFT frameworks in the literature. While the results align with trends reported in related studies, a systematic head-to-head benchmarking would provide stronger evidence of the relative merits of the implemented strategies. Furthermore, the reported results were obtained from validation sets used for model selection, and further evaluation on independent test cohorts would be beneficial to assess generalizability. Full fine-tuning, with all pre-trained parameters updated from the start of adaptation, was not evaluated as a separate comparator.

Third, the sensitivity analysis of the LoRA rank hyperparameter r, applied to convolutional and transposed convolutional layers, is limited to the values tested in the current scenarios. A more extensive exploration of a wider range of architectures and domain shifts is necessary to provide definitive guidance on rank selection.

These limitations highlight key directions for Dante's future development, which will include integrating additional datasets, evaluating various fine-tuning configurations, and exploring unsupervised pretraining techniques to enhance model fitting speed, especially in the absence of segmentation labels for model's pretraining, a common situation in clinical context.

In conclusion, Dante places itself as a potentially valuable tool in the Dafne ecosystem, for the simplified pretraining of models which can be achieved in a clinical research setting, where data is potentially scarce, and which might benefit from a simplified user interface without the requirement of custom software implementation.

| Source Dataset | Fine-Tuning Strategy | Best Avg Dice | Gain vs Scratch | Epoch @85% |
|---|---|---|---|---|
| None (Scratch) | Baseline | 0.7848 | - | 41 |
| CHAOS | LoRA (r=32) | 0.8864 | +10.17 | 11 |
| NIH Dataset | LoRA (r=32) | 0.9568 | +17.20 | 15 |
| NIH Dataset | GU | 0.9555 | +17.08 | 14 |
| CHAOS | GU | 0.8732 | +8.85 | 6 |

**Table 4**: This comparison examines training from scratch versus Dante-based fine-tuning, utilizing CHAOS and NIH pre-training sources. The table illustrates how parameter-efficient fine-tuning, combined with extensive anatomical pre-training, enhances high-accuracy segmentation while requiring minimal local data.

| Scenario | Task | Source Dataset (Pre-training) | Target Dataset (Fine-tuning) | Train/Validation Patients | Source and Target Labels (organs, lesions, …) |
|---|---|---|---|---|---|
| Multi-Organ Transfer (A) | Abdominal MRI | AMOS | CHAOS | 34/8 (Pre-training)<br>16/4 (Fine-tuning) | 15 AMOS abdominal organs * (Source)<br>4 CHAOS abdominal organs ** (Target) |
| Multi-Organ Transfer (B) | Abdominal MRI | CHAOS | AMOS | 16/4 (Pre-training)<br>8/2 (Fine-tuning) | 4 CHAOS abdominal organs * (Source)<br>4 abdominal organs (CHAOS labels) * (Target) |
| Multi-Organ Transfer (C) | Abdominal MRI | NIH Dataset | AMOS | 156/39 (Pre-training)<br>8/2 (Fine-tuning) | 7 NIH labels from a total of 62*** (Source)<br>4 abdominal organs (CHAOS labels) ** |
| Multi-Organ Transfer (D) | Abdominal MRI | NIH Dataset | AMOS | 156/39 (Pre-training)<br>8/2 (Fine-tuning) | 4 NIH labels from a total of 62****<br>4 abdominal organs (CHAOS labels) |
| MS Brain Lesion Adaptation | Brain FLAIR MRI | MSLesSeg | ISBI 2015 | 60/15 (Pre-training)<br>3/2 (Fine-tuning) | Multiple Sclerosis lesion (Source and Target) |

**Table 3**: Experimental Design and Label Mapping Strategy**.** A summary of the pre-training sources, target adaptation tasks, and the anatomical structures segmented in each experimental condition. *AMOS labels: Liver, Right Kidney, Left Kidney, Spleen, Pancreas, Aorta, Inferior Vena Cava, Right Adrenal Gland, Left Adrenal Gland, Gallbladder, Esophagus, Stomach, Duodenum, Bladder, Prostate/Uterus. **CHAOS labels: liver, left kidney, right kidney, spleen. *** NIH labels: liver, left kidney, right kidney, spleen, stomach, inferior vena cava, pancreas. **** NIH labels: stomach, aorta, inferior vena cava, pancreas.

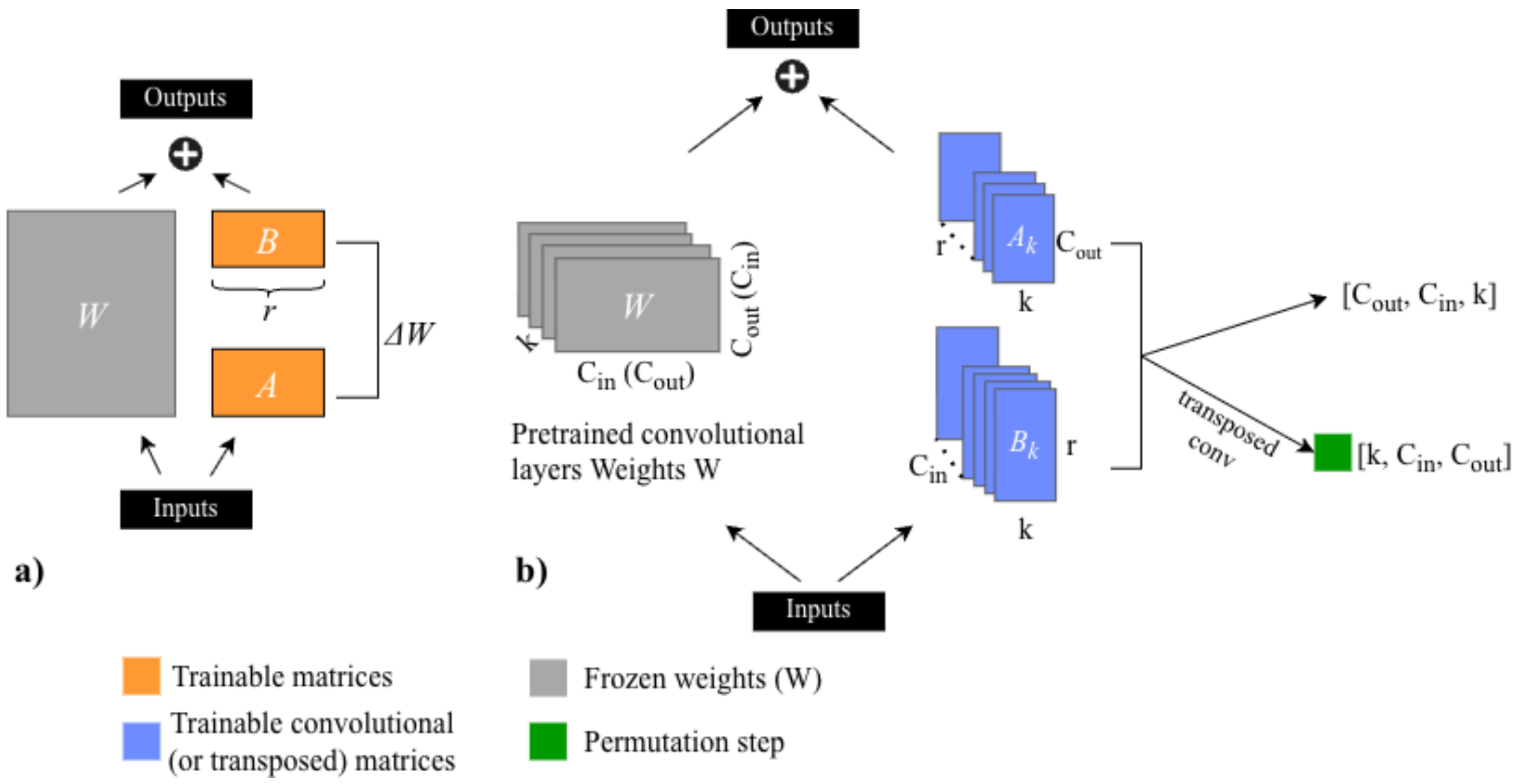

**Figure 2**: LoRA Adaptation for Convolutional Layers in Dante. a) Standard LoRA for linear layers: the weight update ΔW is decomposed into two low-rank trainable matrices, B and A, with rank r, while the original weights W remain frozen and in b) the approach that extends to N-dimensional convolutional layers, where ΔWk is calculated as the product of Bk (k × Cout × r) and Ak (k × r × Cin), leveraging the linearity of the convolution operator. For transposed convolutional layers, a permutation step rearranges the axes of ΔWk to match the expected weight configuration. In both cases, only the low-rank matrices are updated during training.

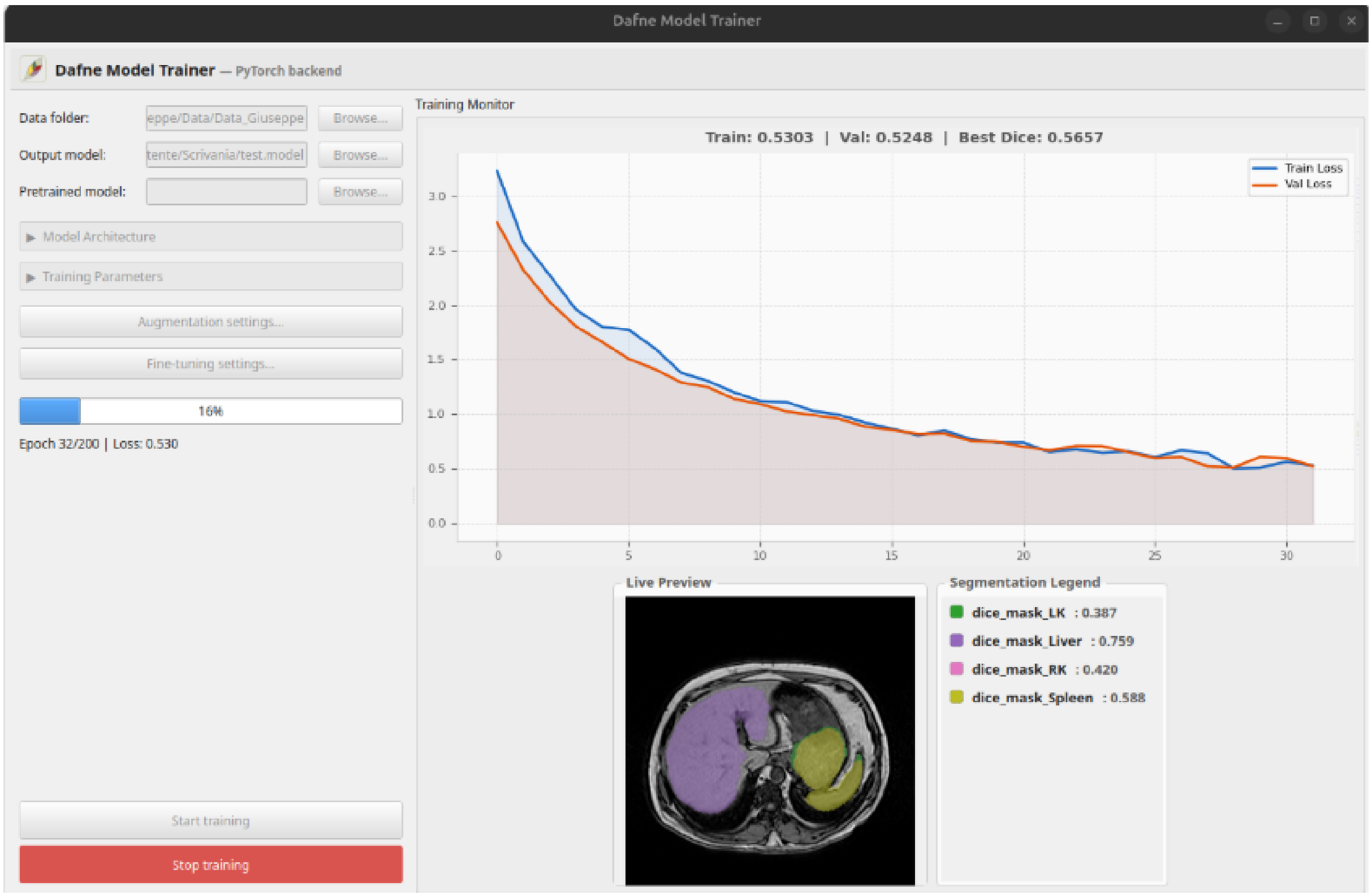

**Figure 3:** Dafne Model Trainer GUI - an example of training from scratch using the CHAOS dataset. During both training from scratch and fine-tuning, the GUI displays validation loss curves and additional information, such as the best average dice score and the current epoch's dice score for each segmented organ.

# Acknowledgements

The authors would like to acknowledge the NIH Clinical Center for providing the abdominal MRI dataset used in this study. The authors also acknowledge the creators and maintainers of the publicly available AMOS, CHAOS, MSLesSeg, and ISBI 2015 datasets, whose contributions made the experimental validation of Dante possible.

# Data availability

The datasets analysed during the current study are available from their original publicly accessible repositories. The AMOS22 dataset is available through Zenodo at https://doi.org/10.5281/zenodo.7262581, and the CHAOS dataset is available through Zenodo at https://doi.org/10.5281/zenodo.3362845. The NIH MRISegmenter dataset, provided by the NIH Clinical Center, is available at https://nihcc.box.com/s/zbrocs18g9ctrl0gb4n3hq7dlotpuwg5. The MSLesSeg dataset is available through Figshare at https://doi.org/10.6084/m9.figshare.27919209. The ISBI 2015 Longitudinal Multiple Sclerosis Lesion Segmentation dataset is described on the official Johns Hopkins University challenge website (https://iacl.ece.jhu.edu/index.php/MSChallenge/data). An alternative publicly accessible copy of the dataset is available on Kaggle (https://www.kaggle.com/datasets/marwa96/isbi-ms-dataset), subject to the access conditions of the respective platform. All datasets were obtained from their original sources and used for the experimental evaluation described in this manuscript.

# Code availability

The Dante source code is available at https://github.com/dafne-imaging/dafne-torch-trainer. The version used in this study is archived at https://doi.org/10.5281/zenodo.24134351

# Funding Declaration

This research did not receive any specific funding from public, commercial, or not-for-profit funding agencies. The study was conducted without dedicated financial support, and no external grants or financial contributions were received specifically for the development, experimental validation, or preparation of this manuscript.

# SUPPLEMENTARY MATERIALS

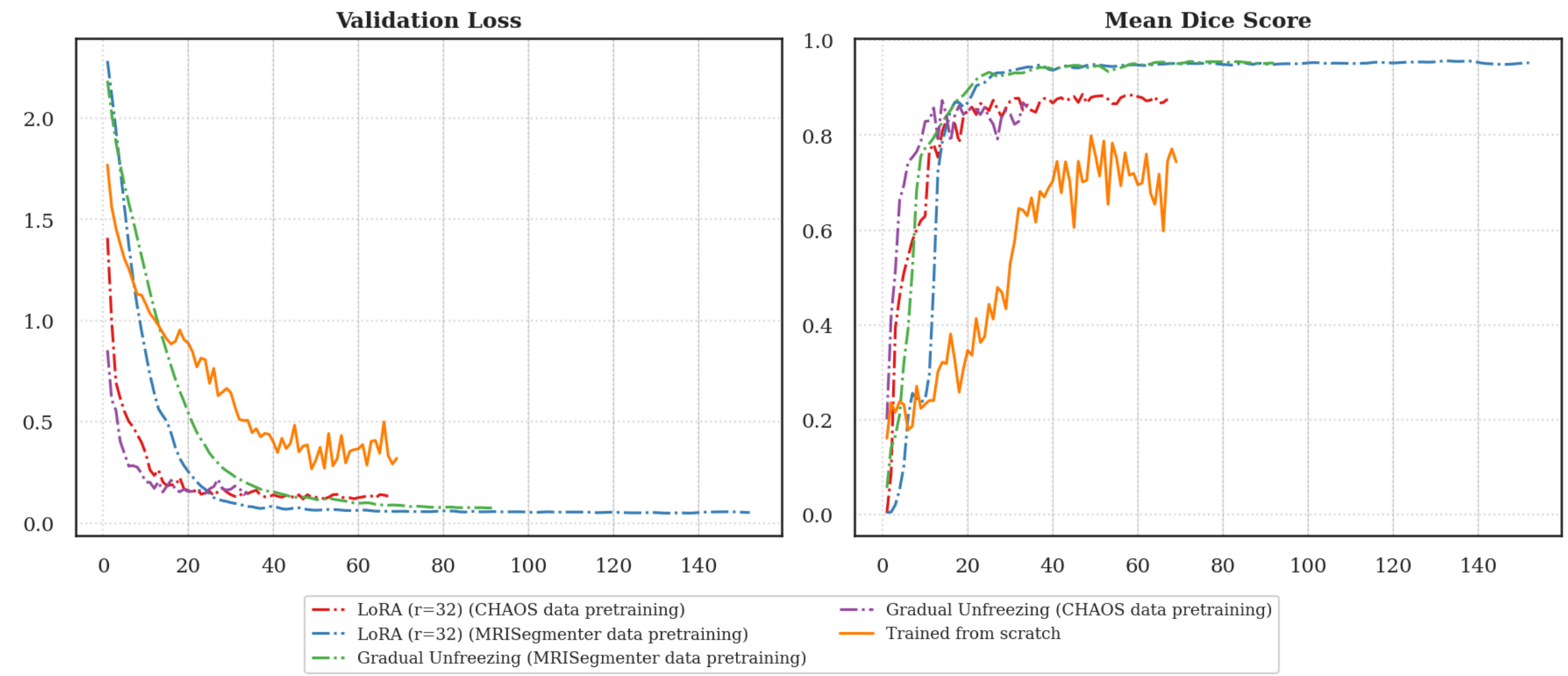

**Figure S1.** Validation loss and DICE curves in the AMOS subset training. The loss and DICE curves on the validation dataset indicate that the baseline model struggles to generalize on the small validation set. In contrast, GU and LoRA (r=32), leveraging prior knowledge of anatomical structures, demonstrate better generalization on the validation dataset.

# SUPPLEMENTARY MATERIALS

| Strategy | Avg. DSC | Liver | Spleen | L. Kidney | R. Kidney | Peak Epoch | Epoch @85% |
|---|---|---|---|---|---|---|---|
| Baseline (Scratch) | 0.8400 | 0.8308 | 0.8213 | 0.8784 | 0.8740 | 171 | 44 |
| GU | 0.8573 | 0.9150 | 0.8329 | 0.8791 | 0.8528 | 69 | 16 |
| LoRA (r=8) | 0.8711 | 0.9225 | 0.8630 | 0.8838 | 0.8333 | 74 | 43 |
| LoRA (r=16) | 0.8726 | 0.9243 | 0.8537 | 0.8849 | 0.8450 | 64 | 35 |
| LoRA (r=32) | 0.8695 | 0.9261 | 0.8553 | 0.8843 | 0.8429 | 93 | 29 |

**Table S1.** Performance comparison for fine-tuning on CHAOS' dataset for average results and individual organs (Liver, Spleen, Left Kidney, and Right Kidney), including the average Dice peak and the epoch required to reach 85% of that peak. GU: Gradual Unfreezing

| Strategy | Avg. DSC | Peak Epoch | Epoch @85% |
|---|---|---|---|
| Baseline (Scratch) | 0.8184 | 61 | 17 |
| GU | 0.8493 | 32 | 1 |
| LoRA (r=16) | 0.8397 | 6 | 2 |
| LoRA (r=32) | 0.8481 | 5 | 2 |

**Table S2:** Performance comparison on the MS target dataset: Dante's fine-tuning techniques generalize better to the target dataset and achieve Dice scores on MS lesions more quickly.